\documentclass{emulateapj}
\usepackage{apjfonts}
\usepackage{graphicx}
\usepackage{amsmath}
\usepackage{natbib}
\usepackage{epsfig}
\newcommand{\etal}{{\it et al.~}}
\def\v#1{{\bf #1}}

\begin{document}

 \title{On Diurnal and Annual Variations of Directional Detection Rates
of Dark Matter}

 \author{Abhijit Bandyopadhyay}
\affil{Department of Physics, RKM Vivekananda University, Belur Math, Howrah 711202, India}
\email{abhi.vu@gmail.com}

 \author{Debasish Majumdar}
\affil{Astroparticle Physics and Cosmology Division, Saha Institute of Nuclear Physics, 1/AF Bidhannagar, Kolkata 700064, India}
\email{debasish.majumdar@saha.ac.in}
 
\thispagestyle{empty}

\sloppy

\begin{abstract}
Direction sensitive direct detection of Weakly Interacting 
Massive Particles (WIMPs) as dark matter would
provide an unambiguous non-gravitational signature of dark matter (DM). 
The diurnal variation of DM signal due to earth's rotation 
around its own axis can be a significant signature for galactic WIMPs.  
Because of particular orientation of earth's axis of rotation
with respect to WIMP wind direction, the apparent direction
of WIMP wind as observed at a detector can alter widely over a day. 
In this work we calculate  the directional detection rates 
with their daily and yearly modulations
in  earth-bound dark matter experiments considering
detailed features of the  geometry and dynamics of the earth-sun system
along with the solar motion in galactic frame.
A separate halo model namely the dark disc model other than the
usual standard halo model for dark matter halo is also considered 
and the results for two models are compared.
We demonstrate the results for two types of
gas detectors namely DRIFT (target material CS2) and NEWAGE 
(target material CF4) that use Time
Projection Chamber techniques for measuring directionality of the recoil 
nucleus. The WIMP mass and recoil energy dependence of the
daily variation of event rates are computed for specific detector
and the sensitive ranges of  mass and recoil energies 
for the considered detector are probed.
\end{abstract}
 

\maketitle

\section{Introduction}
Dark matter is a non-luminous, non-baryonic, pervasive fluid extending 
in and around the bound astronomical systems like galaxy and galaxy
clusters.
This hypothetical proposition emerged from an attempt to reconcile astronomical
observations with Newtonian dynamics. Indirect indications of existence of
dark matter is perceived through their putative gravitational interactions
in several cosmological observations like rotation curves of spiral galaxies,
the gravitational micro-lensing, observations on Virgo and Coma clusters 
\citep{McLaughlin1,Lokas1}
bullet clusters  \citep{Bradac1} etc.
These observations indicate the existence of roughly spherical halo of dark matter
around the visible galaxies.
Most of the constituents of dark matter is believed to be non-relativistic 
(cold dark matter) with Weakly Interacting Massive
Particles (WIMPs) as the plausible candidates. 
The direct detection of the dark matter in terrestrial detectors involves the scattering of the 
target nuclei in a detector when  dark matter interacts with them.
The resulting recoil nuclei deposit their energy in the detector to produce dark matter signal in the detector. 
Dedicated detectors are operational in search of any non-gravitational interaction
of dark matter with ordinary matter to provide 
more insight into the nature of dark matter and its interactions.
The dynamics of rotation of the Milky Way galactic disc
through the halo of dark matter causes 
the earth to experience a wind of WIMPs apparently
flowing along a  direction opposite to that of the motion of the solar system
relative to the dark matter halo.
There are several ongoing/proposed experimental efforts
\citep{CDMS,DAMAa,DAMAb,XENONa,XENONb,CRESSTa,CRESSTb,
CRESSTc,CRESSTd,DRIFTa,DRIFTb,DRIFTc,EDELWEISSa,EDELWEISSb,
EDELWEISSc,NEWAGEa,NEWAGEb,ArDM, 
PICASSO,MIMACa,MIMACb,MIMACc,MIMACd,ZEPLINa,ZEPLINb,
ROSEBUD1,ROSEBUD2,ROSEBUD3,ROSEBUD4,
DMTPCb,DMTPCc}
to directly detect these galactic WIMPs by measuring the 
energy deposition by the WIMPs through their 
scattering off the detector nuclei. Unambiguous identification
of a WIMP signal is severely challenged by presence of
similar background signatures.\\

An annual modulation is expected \citep{drukier,freese} 
in the  detection rates of WIMPs
in the direct detection experiments. This seasonal variation  is an effect of 
earth's revolution around the sun in course of which
the component of earth's orbital velocity parallel to the WIMP flow direction
annually varies causing the earth to encounter different WIMP fluxes
at different times of the year. But the relative motion between the sun and the 
earth is much slower compared to the motion of solar system in galactic halo.
Owing to this, the annual modulation is expected to be
only a few percent of the WIMP detection 
rate and it is extremely difficult
 to extricate it  as a positive signature of dark matter
from seasonal variation of background  rates.
On the contrary, observation of directional anisotropy of the WIMP wind on earth
is a signature which can hardly be mimicked by
any other backgrounds and is potentially more powerful in providing 
unambiguous signature of galactic WIMPs. There are dedicated experiments 
with gas detectors with a Time Projection Chamber (TPC) (DRIFT \citep{DRIFTa,DRIFTb,DRIFTc}, 
NEWAGE \citep{NEWAGEa,NEWAGEb}, MIMAC \citep{MIMACa,MIMACb,MIMACc,MIMACd}, DMTPC \citep{DMTPCb,DMTPCc}) 
or organic crystals 
\citep{CRYSTAL} as detectors, aiming to measure the directional (angular) 
distribution of WIMP induced nuclear recoils. 
The technique involved in the directional measurements 
is the 3-D or 2-D reconstruction of nuclear recoil tracks
with a good spatial resolution followed by the  determination of
sense of the recoil direction (discrimination between
the head and tail of the recoil track). The directional distribution of
the nuclear recoils which carries the imprints of incoming WIMP directions
can be exploited to disentangle  WIMP signals from backgrounds.
\footnote{For a 
detailed description of directional detection
of dark matter see \citep{Sciolla}.}. There is a plethora of papers
where comprehensive studies of various aspects of directional detection of
dark matter has been performed \citep{ddalla,ddallc,ddalld,ddalle,ddallf,ddallg,ddallh,ddalli,ddallj,
GONDOLO1,ddmodela,ddmodelb,ddmodelc,ddmodeld,ddmodele,ddmodelf,
ddmodelg,ddmodelh,ddmodeli,ddmodelj,ddmodelk,ddmodell}.
However, because of earth's rotation about its axis which has a particular
orientation relative to the direction of incoming WIMPs,
the average direction of the WIMP wind with respect to an 
observer fixed on earth,
changes over a sidereal day. Such modulations in the apparent WIMP 
direction leads to daily fluctuations in the directional (angular) 
distribution of WIMP signal observed at a static detector on earth. 
Observation of such diurnal variations in the
directional distribution of the WIMP signal would be a robust
signature for WIMPs   \citep{Spergel:1987kx, collar}.
The directional dependence, however, is  sensitive to
the velocity distribution of WIMPs in the halo of dark matter.
Generally the dark matter halo is considered to be spherical and isothermal 
\citep{halomodel1,halomodel2} with the velocity distribution of dark matter WIMPs with respect to the 
galactic rest frame being Maxwellian which is a steady state solution of the 
collisionless Boltzmann equation \citep{pijushpani}.
Halo model dependence of the directional 
distribution has been  extensively discussed in 
\citep{GONDOLO1}.\\

The directional sensitivity is usually realised and presented
in terms of galactic coordinates where the description is
independent of daily time. In order to comprehend
the manifestation of directional sensitivity in the WIMP signal as registered in
the earth-bound detectors, 
it is essential to describe the directional dependence of 
recoil rates with respect to laboratory (detector) fixed coordinate system.
The transformation from galactic frame to laboratory frame naturally
invokes temporal variation in the directional description of WIMP rates over a 
sidereal day. In this work we explicitly calculate the 
transformation between galactic and laboratory reference frames
taking into account  details of the earth-sun geometry and dynamics and
compute the time dependence of the directional recoil rates
in  direction sensitive terrestrial dark matter experiments.
In order to demonstrate the results of the directional rates calculations
in the laboratory frame we assume the standard halo model
of dark matter with a Maxwellian velocity distribution for the WIMPs.\\

In a previous work \citep{verga} also the authors have studied
the diurnal variation of the direct detection of dark matter
in directional experiments. In  \citep{verga},
they express the WIMP velocity in a coordinate system
where the polar axis ($Z$) is in the direction
of recoiling nucleus. In the present work, however,
the detector fixed coordinate systems are used
and the directional detection rates are explicitly computed
for two specific detectors. In this case the direction
is specified by the direction of recoil 
nucleus with respect to  this detector fixed system.
For the study of diurnal variation in \citep{verga}
the orientation of a detector is  specified in the local frame
which is given by a specific point in the sky designated
by the right ascension $\alpha$ and declination
$\delta$. On the contrary,
in the present work we explicitly show, for specific detectors, the 
actual variation of the WIMP detection rates over a day
for different recoil directions (of the target nucleus)
with respect to the detector fixed frame of reference.  \\

It is also worth investigating the 
impact  if other halo model different from the usual standard
halo model is used for these calculations. On the other hand,
such investigations may be useful to probe the structure of
dark matter halo from the results obtained in actual experiments.
In this paper we consider another halo model namely dark disc model
 \citep{ddisc} and compare the results with those obtained for standard galactic halo model. \\
  
The paper is organised as follows.
In Sec.\ \ref{sec:rates} we present the expressions required to 
calculate the directional recoil rates. Sec.\ \ref{sec:vw}
describes the analytic calculations for obtaining the
results in laboratory frame of reference. 
In  Sec.\ \ref{sec:da} we compute the diurnal and annual variation
of directional detection rates for two terrestrial dark matter experiments.
The response of the detector to the daily variation of event rates
for different ranges of  WIMP masses and  recoil 
energies is computed in this section. 
Finally in Sec.\ \ref{sec:con} we give some concluding remarks.

\section{Angular dependence of the direct detection rates}
\label{sec:rates}
\begin{figure}[h]
\begin{center}
\includegraphics[width=5cm, height=5cm, angle=0]{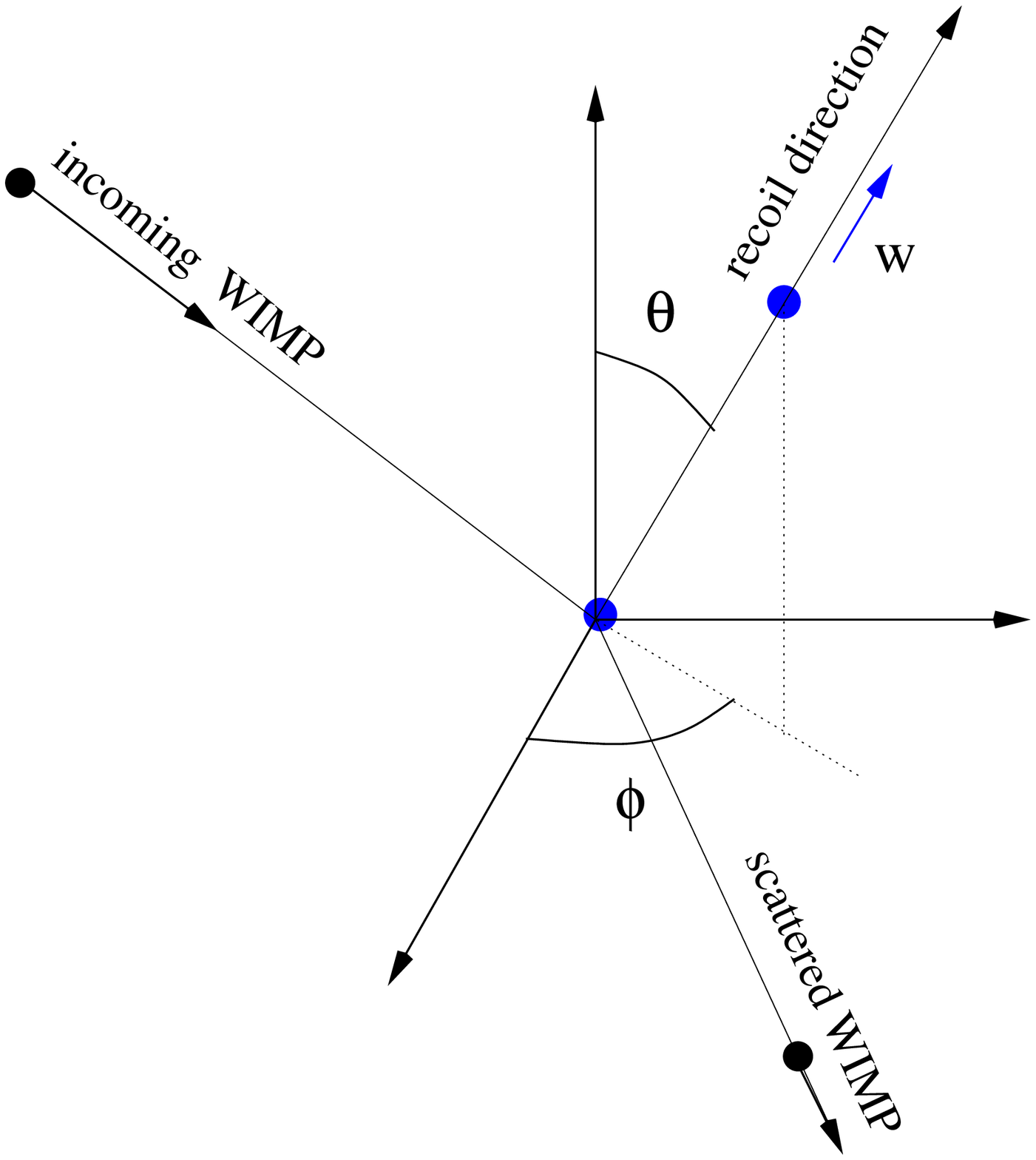}
\caption{\label{fig:f0} Specification of direction of nuclear recoil
in WIMP-nucleus scattering.}
\end{center}
\end{figure}

The differential rate for WIMP induced nuclear recoils per unit detector mass
is given by \citep{GONDOLO1}
\begin{eqnarray}
\frac{dR}{dE}
&=&
\sum_n \frac{\rho}{2\mu_n^2m}C_n\sigma_n(E)
{\cal E}(E) \int_{v>v_n} \frac{f({\bf v})}{v} d^3v
\label{eqdiff1}
\end{eqnarray}

%
where, the summation index $n$ represents a particular nuclear species
of the target nucleus of the detector (e.g. 
$n=C,S$ for $CS_2$ target in DRIFT detector and $n=C,F$
for $CF_4$ target in NEWAGE detector). 
$C_n$ gives the mass fraction of nuclear species $n$ in
the target material. In the above
$\sigma_n(E) = E_{\rm max} (d\sigma_n/dE)$,
($d\sigma_n/dE$) being the differential cross section of
WIMP-nucleus($n$) scattering and $E_{\rm max}$ is the maximum possible
energy transfered between a WIMP and a nucleus of species $n$, which can
be expressed in terms of WIMP speed ($v$), nuclear mass ($M_n$)
and WIMP mass ($m$) as
$E_{\rm max} = 2\mu_n^2 v^2/M_n$, with $\mu_n = mM_n/(m + M_n)$.
In this work 
the local dark matter density $\rho$ 
is taken to be $0.3$ GeV~cm$^{-3}$ and
the detection efficiency (${\cal E}(E)$ for recoil energy $E$) 
to be 100\%. 
Following \citep{GONDOLO1} (and references therein), 
 the Radon transformation of 
the velocity distribution $f(\v{v})$ enables one to write
the differential rate for WIMP induced nuclear recoils 
per unit solid angle per unit detector mass as
\begin{eqnarray}
\frac{dR}{dE\,d\cos\theta \,d\phi} 
&=&
\sum_n \frac{\rho}{4\pi\mu_n^2m}C_n\hat{f}_n(v_n,{\bf w})\sigma_n(E)
{\cal E}(E)
\label{eqdiff}
\end{eqnarray}
The recoil momentum spectrum for nucleus $n$ is
denoted by $\hat{f}_n(v_n,{\bf w})$ where
${\bf w}$ is a unit vector along the direction of
nuclear recoil and $v_n$ ($= c\sqrt{M_nE/2\mu_n^2}$)
represents the minimum velocity of WIMP
required to transfer an amount of energy $E$ to the nucleus $n$ ($c$ is
the speed of light). The function 
$\hat{f}_n(v_n,{\bf w})$ is the 3D Radon transform of 
the velocity distribution $f(\v{v})$ and is given by
$\hat{f}_n(v_n,{\bf w}) = \int \delta(\v{v}\cdot \v{w} - v_n)f(\v{v})d^3v$ \citep{GONDOLO1}.
For a  Maxwellian velocity distribution with velocity 
dispersion $\sigma_v$ for the WIMPs,  
the recoil momentum spectrum in the laboratory
rest frame is given by
\begin{eqnarray}
\hat{f}_n(v_n,{\bf w}) &=& \frac{1}{(2\pi\sigma^2_v)^{1/2}} 
\exp \left [ - \frac{(v_n - {\bf w} \cdot {\bf V})^2} 
{2\sigma_v^2} \right ] 
\label{eqmomspec}
\end{eqnarray}
where, {\bf V} is the average velocity of the WIMPs with 
respect to the detector. The direction of
nuclear recoil in three dimension is described in a chosen frame
of reference by two angles $\theta$ and $\phi$ 
as shown in Fig.\ \ref{fig:f0}. 
The recoil energy-integrated directional differential rate is then given by
\begin{eqnarray}
\frac{dR}{d\cos\theta \ d\phi} 
&=&
\int_{E_{\rm th}} dE \left(\frac{dR}{dE \ d\cos\theta \ d\phi} \right)
\label{eqdedomega}
\end{eqnarray}
$E_{\rm th}$ being the detector threshold energy.  \\

The directional dependence of
the observed recoil rates convolves the WIMP velocity distribution in the 
dark matter halo and average velocity of the WIMPs with respect to the
detector. 
The direction sensitivity of recoil rates are described here,
in terms of the two angles $\theta$ and $\phi$, defined 
in the rest frame of the detector. 
Hence our computation of expression in Eq.\ \ref{eqdedomega} 
directly gives the angular dependence of rates as actually expected 
to be observed in a terrestrial detector.
In a detector-fixed frame of reference  
the velocity of the WIMP wind acquires a time-dependent description 
due to earth's revolution around the sun and earth's rotation about its own
axis. As a result the  observed angular dependence of the recoil rates,
described in the frame of the terrestrial detector,
would exhibit both annual and diurnal variations.
We compute such annual and diurnal variations of the angular
distribution of the event rates in direction sensitive dark matter detectors.

\section{Calculation of $\v{w}\cdot\v{V}$}
\label{sec:vw}
\begin{figure}[h]
\begin{center}
\includegraphics[width=8cm, height=5cm, angle=0]{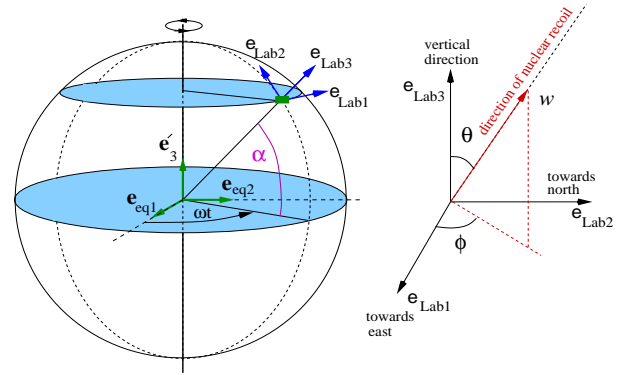}
\caption{\label{fig:lab}Illustration of Laboratory fixed reference frame
$S_{\rm Lab}$ and the direction of nuclear recoil.}
\end{center}
\end{figure}

In this section we present the analytic calculations of the 
direction sensitive term $\v{w}\cdot\v{V}$ (Eq.\ \ref{eqmomspec}) in terms 
of the angles $\theta$ and $\phi$ that specify the recoil direction
$\v{w}$ in a terrestrial dark matter detector. We describe the direction 
sensitivity in a laboratory fixed frame of reference $S_{\rm Lab}$,
denoted in terms of a right-handed system 
($\v{e_{Lab1}}$, $\v{e_{Lab2}}$, $\v{e_{Lab3}}$), 
which are unit vectors pointing respectively towards east, 
north and vertical direction in the sky at the 
detector location. As shown in  Fig.\ \ref{fig:lab},
a particular direction of nuclear recoil, denoted by the unit vector $\v{w}$,
can be expressed in terms of its components in the reference frame  
$S_{\rm Lab}$ as 
\begin{eqnarray}
\v{w} &=& 
\sin\theta\cos\phi \ \v{e_{Lab1}} + 
\sin\theta\sin\phi \ \v{e_{Lab2}} +
\cos\theta \ \v{e_{Lab3}}
\end{eqnarray}
The average velocity $\v{V}$ of WIMP with respect to detector can be written
as \citep{GONDOLO1}
\begin{eqnarray}
\v{V} &=& \v{V_{WG}} - \v{V_{SG}}  - \v{V_{ES}} - \v{V_{LE}}
\label{eq:vform}
\end{eqnarray}
where, $\v{V_{WG}}$ is the velocity of WIMP with respect to 
the galactic centre. In standard halo model $\v{V_{WG}} = 0$.
$\v{V_{SG}}$ is the velocity of sun relative to the galactic centre
which points towards the Cygnus constellation.
$\v{V_{ES}}$ is the velocity of the centre of mass of the
earth relative to the sun and $\v{V_{LE}}$ is the velocity of the laboratory 
(detector) relative to the centre of mass of the earth.
In the following subsections we present explicit
calculations for the quantities
$\v{w}\cdot\v{V_{LE}}$, $\v{w}\cdot\v{V_{ES}}$ and $\v{w}\cdot\v{V_{SG}}$. In obtaining the analytical expressions 
we have neglected the small eccentricity (0.0167) of the Earth's orbit around the sun.

\subsection{Expression for $\v{w}\cdot\v{V_{LE}}$}
The velocity of the detector with respect to the centre of the mass of the
earth is along the direction of $\v{e_{Lab1}}$ and can be written as
\begin{eqnarray}
\v{V_{LE}} 
&=& |\v{V_{LE}}|  \ \v{e_{Lab1}}
= \omega R_\oplus \cos\alpha \ \v{e_{Lab1}}
\end{eqnarray}
$R_\oplus$ being the radius of earth and 
$\omega \ (=2\pi/24 \ {\rm rad / hour})$ is 
the angular speed of earth's rotation about its  axis and $\alpha$
is the latitude of the detector location. Therefore
\begin{eqnarray}
\v{w}\cdot \v{V_{LE}} &=& \omega R_\oplus \cos\alpha\sin\theta\cos\phi
\end{eqnarray}

\subsection{Expression for $\v{w}\cdot\v{V_{ES}}$}
To find $\v{w}\cdot \v{V_{ES}}$ we bring in 
three different frames of references: 
$S_{eq}$, $S_{1\odot}$ and $S_{2\odot}$
which are chosen as follows. The frame 
$S_{eq}$, as shown in Fig.\ \ref{fig:lab}, has a origin  
fixed at the centre of mass of earth and is described by
a right-handed system of orthonormal vectors 
$(\v{e_{eq1}}, \v{e_{eq2}}, \v{e^\prime_3})$.
The plane spanned by the unit vectors $\v{e_{eq1}}$ and $\v{e_{eq2}}$ defines the plane of the equator with
$\v{e_{eq1}}$ pointing vertically upward direction at equator at 12 midnight.
$\v{e^\prime_3}$ is a unit vector pointing towards
direction of angular velocity of earth's rotation and therefore defines the 
earth's axis of rotation. The transformation
between the frames $S_{\rm Lab}$ and $S_{eq}$ frame is given by

\begin{figure}[h]
\includegraphics[width=4.5cm, height=4.5cm, angle=0]{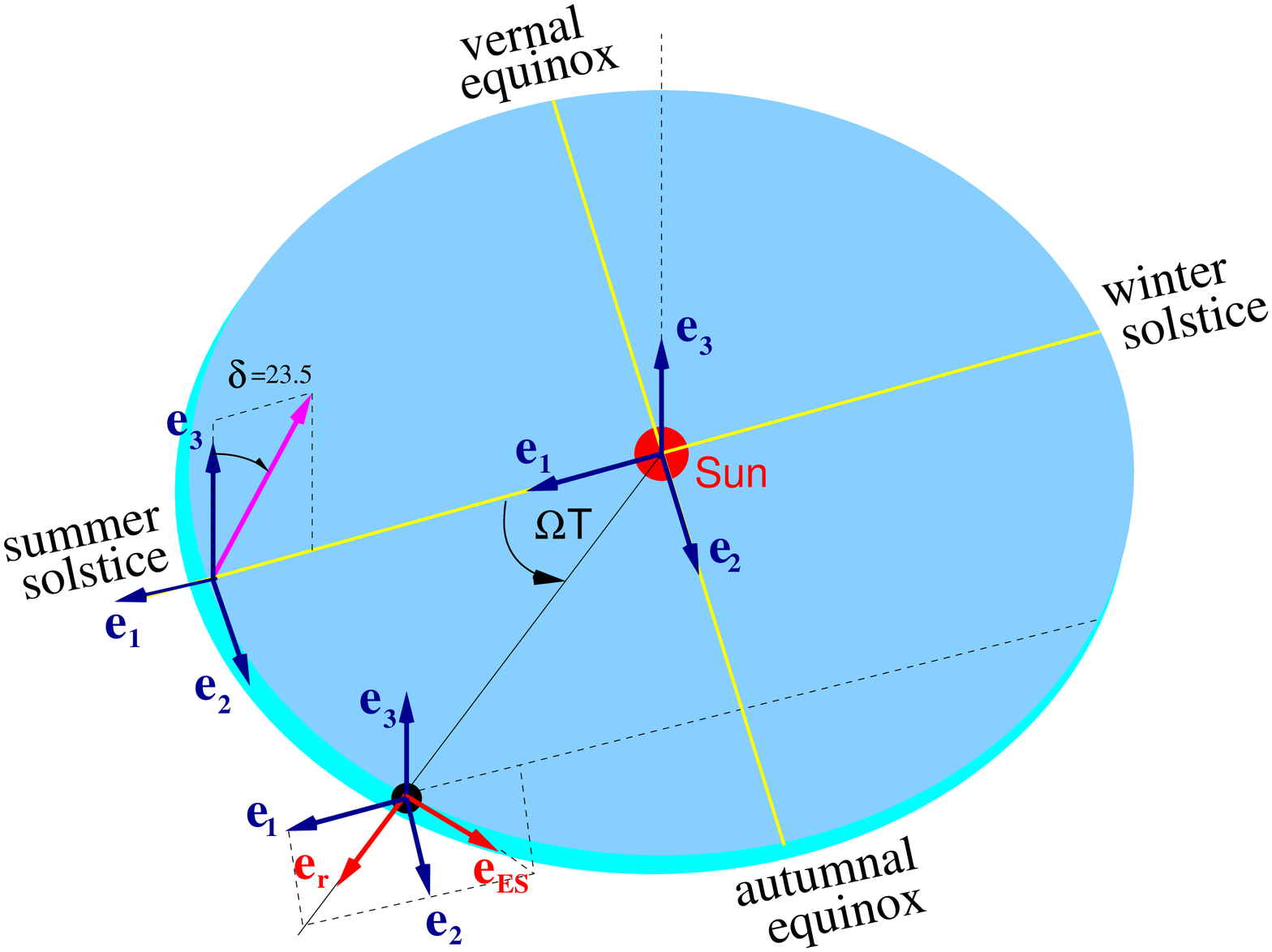}
\vglue -4.5cm \hglue 4.2cm
\includegraphics[width=4.5cm, height=4.3cm, angle=0]{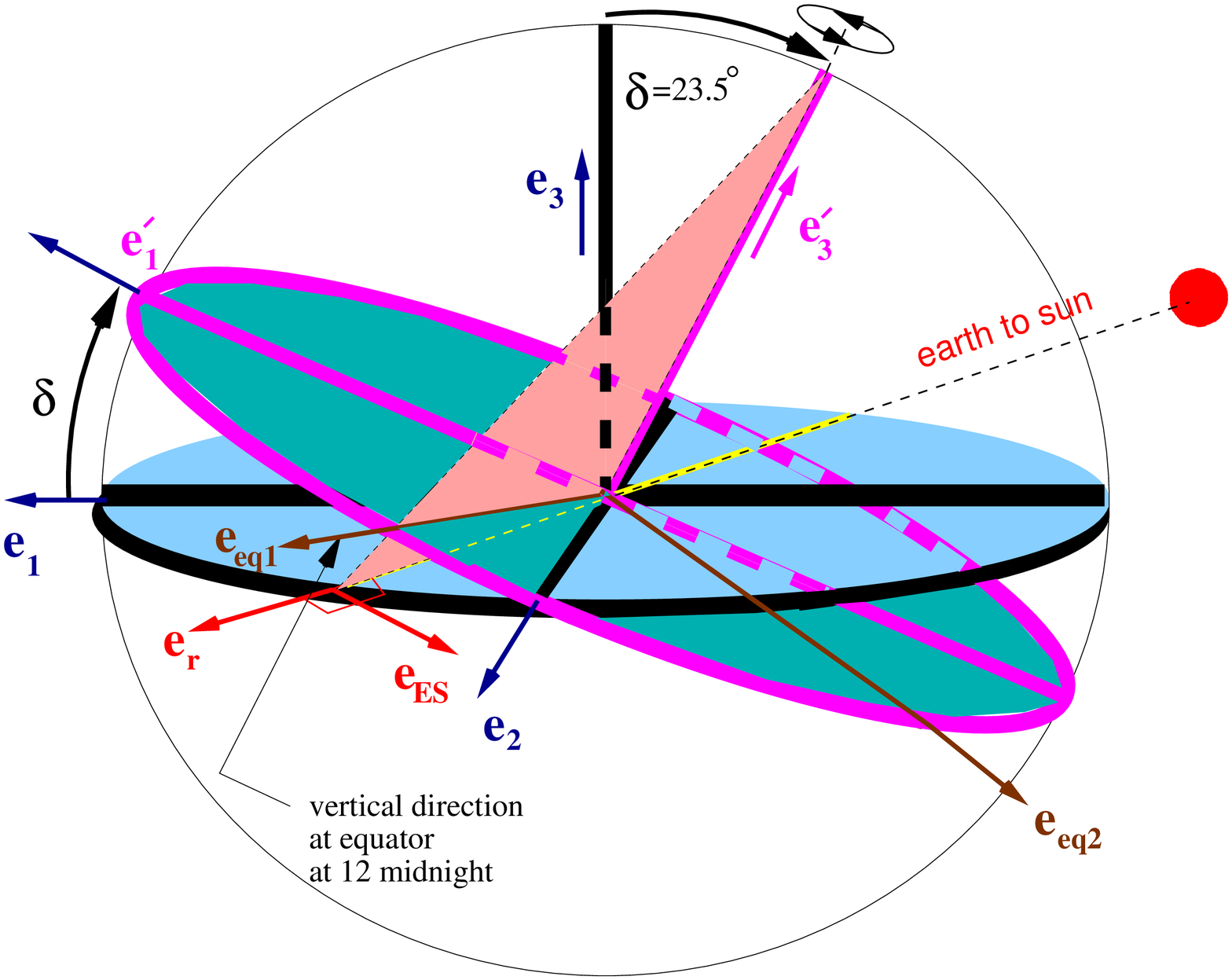}
\caption{\label{fig:sun} Illustration of reference frames
$S_{1\odot}$, $S_{2\odot}$ and $S_{\rm eq}$ (described in text) }
\end{figure}

{\small
\begin{eqnarray}
\begin{pmatrix}\v{e_{eq1}} \cr \v{e_{eq2}} \cr \v{e^\prime_3} \end{pmatrix}
&=&
\begin{pmatrix} - \sin\omega t & -\sin\alpha\cos\omega t & \cos\alpha\cos\omega t\cr
\cos\omega t  & -\sin\alpha\sin\omega t& \cos\alpha\sin\omega t\cr
0 & \cos\alpha & \sin\alpha \end{pmatrix}
\begin{pmatrix} \v{e_{Lab1}} \cr \v{e_{Lab2}} \cr \v{e_{Lab3}} \end{pmatrix}
\label{eq:LEQ}
\end{eqnarray}}
where, $t$ is the time elapsed after 12 midnight.
$S_{1\odot}$ and $S_{2\odot}$ are two frames of reference with their origins
fixed at the sun and are shown in  Fig.\ \ref{fig:sun}. They are described 
respectively by the two right-handed systems
($\v{e_1},\v{e_2},\v{e_3}$) and 
($\v{e^\prime_1},\v{e_2},\v{e^\prime_3}$)
where, $\v{e_1}$, $\v{e_2}$ are mutually perpendicular unit vectors
in the ecliptic plane, with
$\v{e_1}$ pointing towards the centre of mass of earth at 
summer solstice from the sun. $\v{e_3}$ is 
a unit normal to the ecliptic pointing towards the orbital angular
velocity of earth. 
The axis of earth's rotation for all time of the year is parallel 
to $\v{e_1}-\v{e_3}$ plane making an angle $\delta=23.5^o$ with the 
$\v{e_3}$ and at summer  solstice its tilt is towards sun.
The mutually orthogonal unit vectors
$\v{e^\prime_1}$ and $\v{e_2}$ define the plane of the 
equator. The transformation  between frames
$S_{1\odot}$ and $S_{2\odot}$ is then given by
%
%
\begin{eqnarray}
\begin{pmatrix}\v{e_1} \cr \v{e_2} \cr \v{e_3}\end{pmatrix}
&=&
\begin{pmatrix}\cos\delta & 0 & -\sin\delta \cr 
               0 & 1 & 0\cr 
                \sin\delta & 0 & \cos\delta \cr\end{pmatrix}
\begin{pmatrix}\v{e^\prime_1} \cr \v{e_2} \cr \v{e^\prime_3}\end{pmatrix}
\label{eq:S1S2}
\end{eqnarray}
A unit vector along the radius vector of earth 
with respect to the sun at a time $T$ of the year (counted from the day of 
summer solstice) is given by
\begin{eqnarray}
\v{e_r} 
&=& \cos\Omega T \ \v{e_1} + \sin\Omega T \ \v{e_2}\nonumber\\
&=&\cos\Omega T \cos\delta \ \v{e^\prime_1} + \sin\Omega T \ \v{e_2}
- \cos\Omega T \sin\delta \ \v{e^\prime_3}
\end{eqnarray}
where, $\Omega = (2\pi {\rm ~radians/year})$ 
is the orbital angular speed of the earth about sun.
Since $\v{e_{eq1}}$ is along the vertical upward direction at equator at
12 midnight it is parallel to the line of intersection of two planes: the
plane containing the axis of the rotation of earth 
($\v{e^\prime_3}$) and the radius vector joining sun to centre of earth 
and the plane of the equator. 
The normal to the two planes: the plane containing the vectors $\v{e^\prime_3}$ 
\& $\v{e_r}$ and the plane of the equator respectively are 
$\v{e^\prime_3}\times \v{e_r}$ and $\v{e^\prime_3}$. The perpendicular
to both of these normals define the line of intersection of the two planes
and along which the unit vector is
\begin{eqnarray}
\v{e_{eq1}} 
&=&
\frac{(\v{e^\prime_3} \times \v{e_r}) \times\v{e^\prime_3}}
{|(\v{e^\prime_3} \times \v{e_r}) \times\v{e^\prime_3}|} \nonumber\\
&=&
\frac{ \cos\Omega T\cos\delta \ \v{e^\prime_1} + \sin\Omega T \ \v{e_2} }
{\sqrt{\cos^2\Omega T\cos^2\delta +\sin^2\Omega T }}
\end{eqnarray}
The unit vector $\v{e_{eq2}}$ is then obtained as
\begin{eqnarray}
\v{e_{eq2}} &=&  \v{e^\prime_3} \times \v{e_{eq1}} \nonumber\\
&=& \frac{-\sin\Omega T \ \v{e^\prime_1} + \cos\Omega T\cos\delta \ \v{e_2}}
{\sqrt{\cos^2\Omega T\cos^2\delta +\sin^2\Omega T }}
\end{eqnarray}
Therefore the transformation between reference frames $S_{2\odot}$ 
and $S_{eq}$ is given by
\begin{eqnarray}
\begin{pmatrix}\v{e^\prime_1} \cr \v{e_2} \cr \v{e^\prime_3} \end{pmatrix}
&=&
\begin{pmatrix} y\cos\Omega T\cos\delta  & -y\sin\Omega T   & 0 \cr
y\sin\Omega T & y\cos\Omega T\cos\delta & 0 \cr
0 & 0 & 1\end{pmatrix}\begin{pmatrix}\v{e_{eq1}} \cr \v{e_{eq2}} \cr \v{e^\prime_3}\end{pmatrix}
\label{eq:EQS2}
\end{eqnarray}
where 
\begin{eqnarray}
y = \frac{1}{\sqrt{\cos^2\Omega T\cos^2\delta +\sin^2\Omega T }}
\end{eqnarray}
Using Eqs.\ (\ref{eq:LEQ},\ref{eq:S1S2},\ref{eq:EQS2}) we can write the
transformation equation between the frames  $S_{1\odot}$ and $S_{\rm Lab}$ as
\begin{eqnarray}
\begin{pmatrix}\v{e_1} \cr \v{e_2} \cr \v{e_3}\end{pmatrix}
&=&
\begin{pmatrix}\cos\delta & 0 & -\sin\delta \cr 
               0 & 1 & 0\cr 
                \sin\delta & 0 & \cos\delta \end{pmatrix}
\begin{pmatrix} y\cos\Omega T\cos\delta  & -y\sin\Omega T   & 0 \cr
y\sin\Omega T & y\cos\Omega T\cos\delta & 0 \cr
0 & 0 & 1 \end{pmatrix}\nonumber\\
&& \times
\begin{pmatrix} - \sin\omega t & -\sin\alpha\cos\omega t & \cos\alpha\cos\omega t\cr
\cos\omega t  & -\sin\alpha\sin\omega t& \cos\alpha\sin\omega t\cr
0 & \cos\alpha & \sin\alpha \end{pmatrix}
\begin{pmatrix}\v{e_{Lab1}} \cr \v{e_{Lab2}} \cr \v{e_{Lab3}} \end{pmatrix}
\label{eq:fulltranf}
\end{eqnarray}
A unit vector along the direction of instantaneous orbital velocity of
earth with respect to sun at time $T$ of the year is given by
\begin{eqnarray}
\v{e_{ES}} 
&=& -\sin\Omega T \ \v{e_1} + \cos\Omega T \ \v{e_2}
\end{eqnarray}
Using the transformation in Eq.\ \ref{eq:fulltranf} we can express it
in the laboratory frame of reference as 
\begin{eqnarray}
\v{e_{ES}} &=& y\left(\cos\delta\cos\omega t - \frac{1}{2}\sin2\Omega T\sin^2\delta
\sin\omega t\right) \ \v{e_{Lab1}}\nonumber\\
&& 
+\Big{(}\cos\alpha\sin\delta\sin\Omega T 
- y\sin\alpha\cos\delta\sin\omega t  \nonumber\\
&& 
-\frac{y}{2}\sin\alpha\sin2\Omega T\sin^2\delta\cos\omega t 
\Big{)} \ \v{e_{Lab2}}\nonumber\\
&&
+\Big{(}\sin\alpha\sin\delta\sin\Omega T 
+ y\cos\alpha\cos\delta\sin\omega t  \nonumber\\
&&
+\frac{y}{2}\cos\alpha\sin2\Omega T\sin^2\delta\cos\omega t \Big{)} \ \v{e_{Lab3}}
\end{eqnarray}
Finally we get
{\small
\begin{eqnarray}
\v{w} \cdot \v{V_{ES}}
&=& |\v{V_{ES}}| \ (\v{w} \cdot \v{e_{ES}}) \nonumber\\
&=& |\v{V_{ES}}| \bigg{[}
y\left(\cos\delta\cos\omega t - \frac{1}{2}\sin2\Omega T\sin^2\delta
\sin\omega t\right)\sin\theta\cos\phi  \nonumber\\
&+& 
\Big{(}\cos\alpha\sin\delta\sin\Omega T 
- y\sin\alpha\cos\delta\sin\omega t  \nonumber\\
&&
-\frac{y}{2}\sin\alpha\sin2\Omega T\sin^2\delta\cos\omega t 
\Big{)} \sin\theta\sin\phi  \nonumber\\
&+&
\Big{(}\sin\alpha\sin\delta\sin\Omega T 
+ y\cos\alpha\cos\delta\sin\omega t \nonumber\\
&&
+\frac{y}{2}\cos\alpha\sin2\Omega T\sin^2\delta\cos\omega t \Big{)}
\cos\theta \bigg{]} 
\end{eqnarray}}
where, we have taken the magnitude $|\v{V_{ES}}|$ of earth's orbital speed as
$30$ km/s \citep{plankthesis}

\subsection{Expression for $\v{w}\cdot\v{V_{SG}}$}
%
\begin{figure}[h]
\includegraphics[width=4.0cm, height=4cm, angle=0]{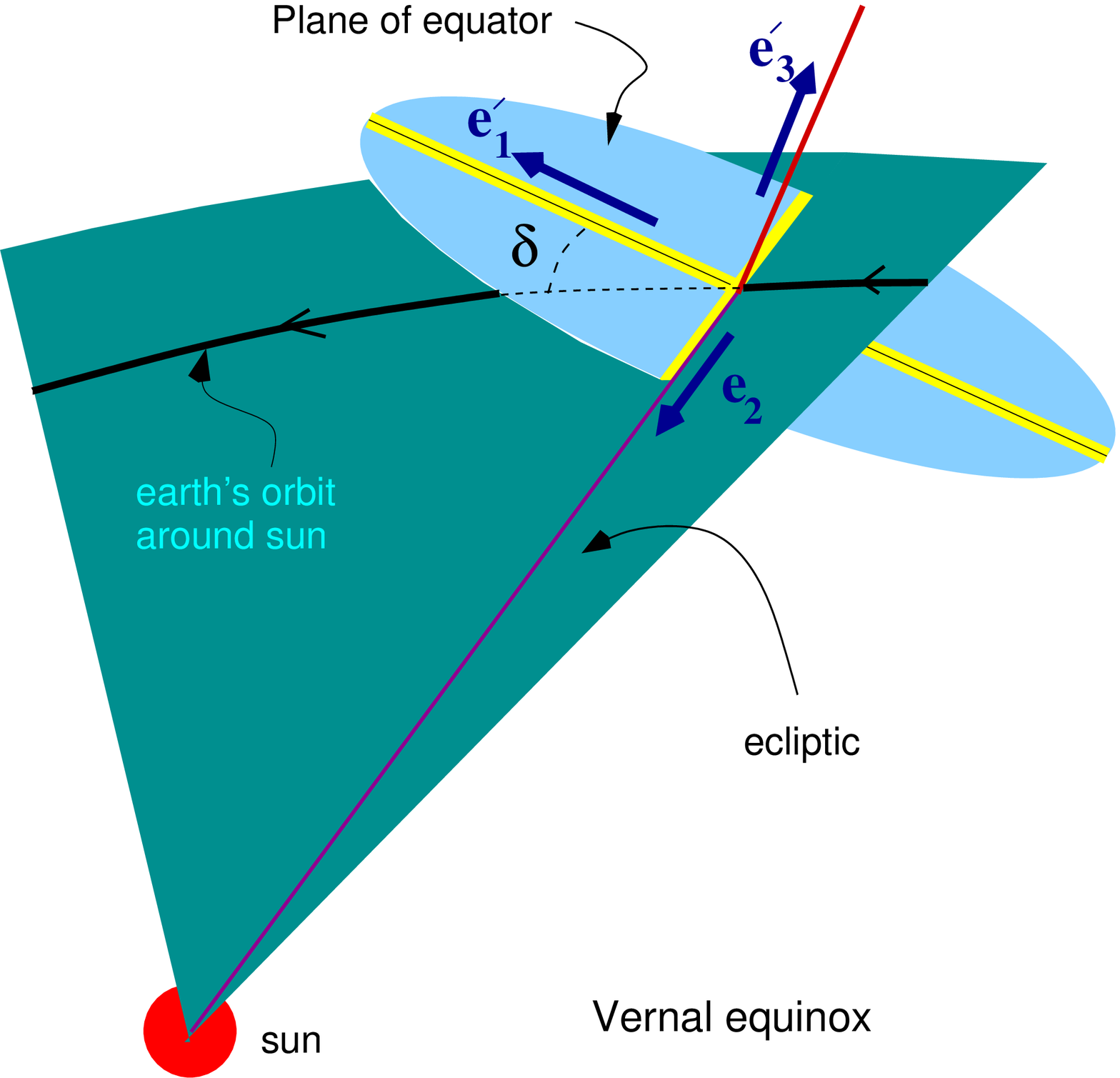}
\vglue -4.0cm \hglue 4cm
\includegraphics[width=4.0cm, height=4cm, angle=0]{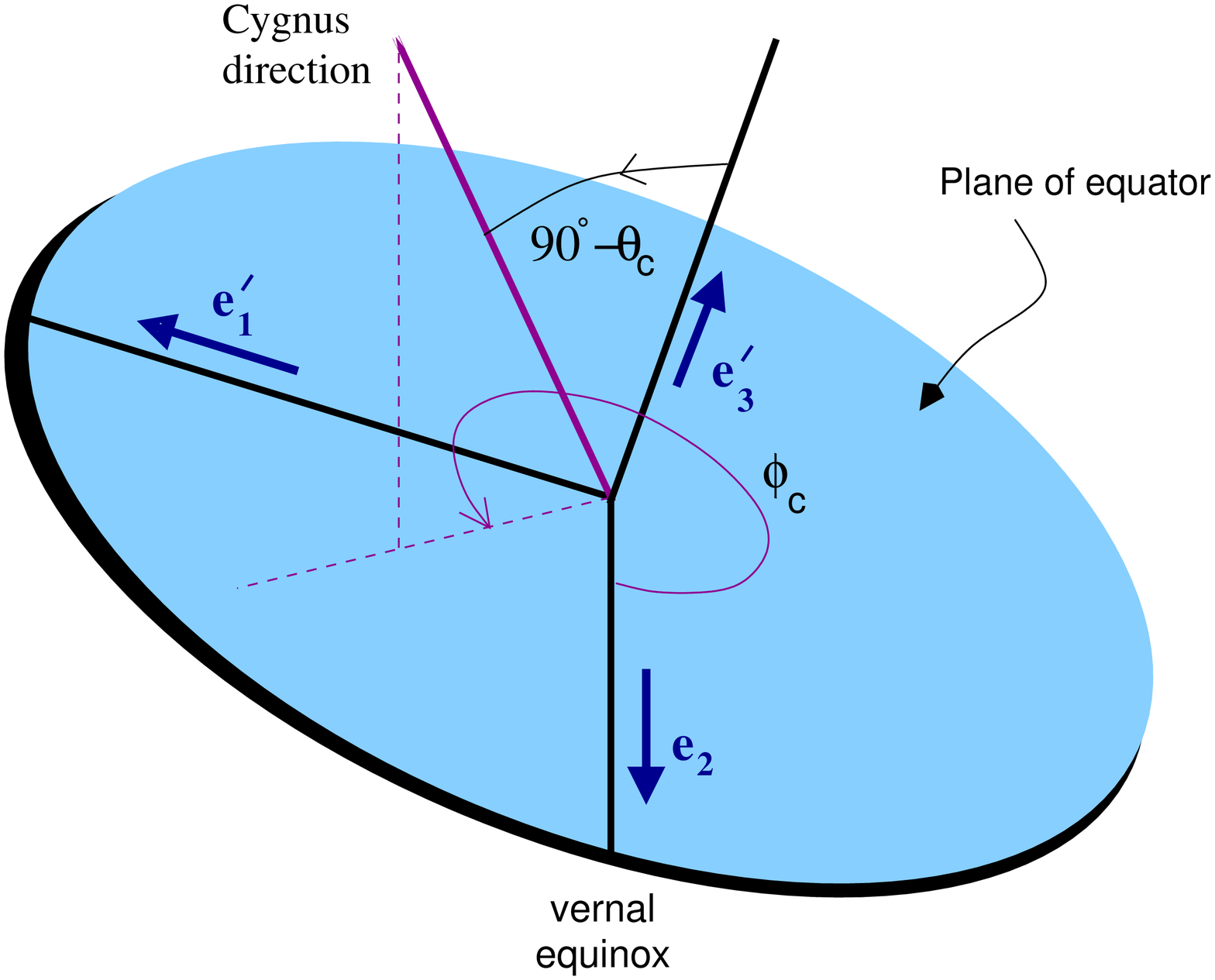}
\caption{\label{fig:gal} Left panel: Relative orientation of
ecliptic and plane of equator at the position of vernal (March) equinox.
Right panel: Declination and right ascension of Cygnus direction or direction
of velocity of sun with respect to galactic centre  has been illustrated.}
\end{figure}
The velocity $\v{V_{SG}}$ of the sun with respect to the galactic centre 
happens to point towards the direction of Cygnus constellation. 
The orientation of Cygnus constellation in celestial coordinate system 
is given by a declination of $\theta_c = 42^\circ$
and a right ascension of $20.62$ hours \citep{cygnus}. This implies that the vector
$\v{V_{SG}}$ makes an angle $90^\circ - \theta_c$ 
with the axis of rotation of earth ($\v{e^\prime_3}$) and its projection
on the plane of the equator subtends an angle 
$\phi_c = 20.62 h\simeq 309^o$ 
when measured  (anticlockwise) 
from the direction in sky where the
sun crosses the celestial equator at vernal equinox.
This  direction is defined by $\v{e_2}$. 
Therefore a unit vector along the direction of
$\v{V_{SG}}$ can be written in the frame
$S_{2\odot} : (\v{e^\prime_1},\v{e_2},\v{e^\prime_3})$ as
\begin{eqnarray}
\v{e_{SG}} &=& -\cos\theta_c\sin\phi_c \ \v{e^\prime_1}
+  \cos\theta_c\cos\phi_c  \ \v{e_2}
+ \sin\theta_c \v{e^\prime_3}
\end{eqnarray}
The transformation matrix between the frames of reference
$S_{2\odot} : (\v{e^\prime_1},\v{e_2},\v{e^\prime_3})$ 
and $S_{\rm Lab} : (\v{e_{\rm Lab1}},\v{e_{\rm Lab2}},\v{e_{\rm Lab3}})$
can be obtained using Eqs.\ (\ref{eq:LEQ},\ref{eq:EQS2})  as
\begin{eqnarray}
\begin{pmatrix}\v{e^\prime_1} \cr \v{e_2} \cr \v{e^\prime_3}\end{pmatrix}
&=&
\begin{pmatrix} y\cos\Omega T\cos\delta  & -y\sin\Omega T   & 0 \cr
y\sin\Omega T & y\cos\Omega T\cos\delta & 0 \cr
0 & 0 & 1 \end{pmatrix} \nonumber\\
&\times &
\begin{pmatrix} - \sin\omega t & -\sin\alpha\cos\omega t & \cos\alpha\cos\omega t\cr
\cos\omega t  & -\sin\alpha\sin\omega t& \cos\alpha\sin\omega t\cr
0 & \cos\alpha & \sin\alpha \end{pmatrix}
\begin{pmatrix}\v{e_{Lab1}} \cr \v{e_{Lab2}} \cr \v{e_{Lab3}} \end{pmatrix} \nonumber\\
\label{eq:halftranf}
\end{eqnarray}
Exploiting the above equations, $\v{e_{SG}}$ can be expressed in terms of
the laboratory frame of reference as
\begin{eqnarray}
\v{e_{SG}} 
&=&
(-A\sin\omega t + B\cos\omega t) \ \v{e_{Lab1}} \nonumber\\
&&+ \Big{[}-\sin\alpha(A\cos\omega t + B\sin\omega T) + \cos\alpha\sin\theta_c\Big{]}
\ \v{e_{Lab2}} \nonumber\\
&&
+\Big{[}(\cos\alpha(A\cos\omega t + B\sin\omega T) + \sin\alpha\sin\theta_c \Big{]}
\ \v{e_{Lab3}}
\end{eqnarray}
where, 
\begin{eqnarray}
A &=& y\cos\theta_c (-\sin\phi_c\cos\Omega T\cos\delta  
+ \cos\phi_c \sin\Omega T )\\
B &=& y\cos\theta_c (\cos\phi_c \cos\Omega T\cos\delta
+\sin\phi_c\sin\Omega T)
\end{eqnarray}
Thus we have
\begin{eqnarray}
\v{w} \cdot \v{V_{SG}} 
&=& 
| \v{V_{SG}}| \ (\v{w} \cdot \v{e_{SG}} )  \nonumber\\
&=& | \v{V_{SG}}| \Bigg{\{}
\sin\theta\cos\phi(-A\sin\omega t + B\cos\omega t) \nonumber\\
&&+ \sin\theta\sin\phi\Big{[}-\sin\alpha(A\cos\omega t + B\sin\omega T) + \cos\alpha\sin\theta_c\Big{]}\nonumber\\
&&+\cos\theta\Big{[}(\cos\alpha(A\cos\omega t + B\sin\omega T) + \sin\alpha\sin\theta_c \Big{]} \Bigg{\}}.
\label{eq:25}
\end{eqnarray}
Although the physical range of $|\v{V_{SG}}|$ is given by 
170 km/sec $ \leq |\v{V_{SG}}| \leq$ 270 km/sec (90 \% C.L.) \citep{dmag1,dmag2,bovy}, in the present work we consider the central
value of  $|\v{V_{SG}}|$ as $220 {\rm ~km/s}$.

\section{Diurnal and annual variations of directional detection rates}
\label{sec:da}
In this section we calculate the diurnal and annual variations
of the direct detection rates with respect to 
the observed direction of recoil nuclei with the formalism 
described in the previous sections. In order to estimate 
such detection rates, the detectors are considered to be direction sensitive, i.e.
the directionality of reconstructed tracks of the recoil nuclei are apparent from the 
observed signals (i.e. the recoil tracks will  have no head-tail ambiguity). 
We also
assume the detectors to be 100\% efficient in the present work. 
From Eq.\ \ref{eqdiff} it is evident that, 
these rates depend on the target materials and location (latitude) 
of the detector. For  
order of magnitude estimations of such rates therefore, the only 
information of detector properties in this calculation are
given in terms of the mass and atomic 
numbers of the target  nuclei
and the mass fraction ($C_n$) of each type of target nucleus (in case of 
diatomic or multiatomic targets), as required to compute Eq.\ \ref{eqdiff}.
In this work, we consider two  directional detectors 
namely DRIFT  (target material CS$_2$) \citep{DRIFTa,DRIFTb,DRIFTc} and NEWAGE 
(target material CF$_4$) \citep{NEWAGEa,NEWAGEb}.\\

The DRIFT detector at Boulby underground laboratory
(54$^\circ$33$^\prime$N,$0^\circ 49^\prime$W) is designed to  
detect WIMPs via nuclear recoils
within the fiducial volume of the detector.
The detector has the ability of not 
only counting nuclear recoils but also determining the 
direction of nuclear recoils
by reconstructing three dimensional tracks of nuclear recoils. 
The DRIFT II detector is a Time Projection Chamber (TPC) 
containing gaseous CS$_2$ at a pressure of 40 torr. 
It consists of two back to back TPCs on either side
of a central photocathode. The ionization electrons resulting from the
scattering of WIMPs off the C or S atoms of CS$_2$ gas get captured by
the electronegative CS$_2$ molecules to form CS$_2^{-}$ ions. These ions 
drift in the electric field towards the anode plate where a multiwire 
proportional chamber consisting of a two dimensional grid of wires are used 
as readouts. The CS$_2$ ions experience only thermal diffusion
allowing a 3D reconstruction of the recoil track range and direction.
The NEWAGE is also a direction sensitive dark matter search experiment 
at Kamioka underground laboratory 
($36^\circ 25^\prime $N, $137^\circ 18^\prime $E).
It uses a three dimensional gaseous tracking detector (Micro-TPC) 
which has Micro Pixel Chamber as its readout. 
The gaseous micro-time projection chamber 
is filled with CF$_4$ at 152 torr and is read by a two dimensional fine 
pitch imaging device called $\mu-$PIC consisting of orthogonally formed
cathode and anode strips. The details of detector fabrication
has been described in \citep{Nishimura:2009zz}. 
These kind of techniques are used to reconstruct the WIMP 
induced nuclear tracks in both two and three-dimensions which then
enable one to determine the directionality of the recoil track.
The recoil energy threshold for the DRIFT experiment can be $\sim$ tens of keV
\citep{plankthesis}. The NEWAGE experimental collaboration has mentioned a 
recoil energy
range $\sim$ 100$-$400 keV for their  WIMP-search analysis. The lower limit of 
this is set by analysing the background gamma rejection energy \citep{surface}.
Since the WIMP induced events will be small it is required to have very low 
threshold for the dark matter detectors albeit it increases the unwanted 
backgroud and unwanted neutron signals. The expected WIMP induced recoil 
energy range is $\sim$ 1 - 50 keV. Shielding etc. are required with specific techniques to reduce the unwanted signals/backgrounds. 
In this work we have presented the results with zero threshold energy
for both DRIFT and NEWAGE.\\

Owing to the apparent directionality of WIMP wind
an anisotropy is expected in distribution of nuclear recoils.
The directional rates are also expected to exhibit a daily modulation
due to rotation of the earth. Such signals of daily modulation can
provide an effective background discrimination and thereby producing
powerful signature of galactic dark matter WIMPs.\\

There are two dominant types of WIMP-nucleus elastic scattering - 
(i) spin-independent interaction, where WIMP interacts with the
nuclear mass and (ii) spin-dependent interaction, where the 
WIMP interaction with the nuclear spin in considered. The WIMP-nucleus
scattering cross section can have contributions from either or both
of the two interactions. In case of DRIFT detector with target material 
CS$_2$, only spin-independent interaction contributes to the total scattering
cross sections as both C and S nuclei have zero spin at the ground state.
On the other hand, since F nucleus has a ground state spin $1/2$,
the NEWAGE detector with target material CF$_4$ is sensitive to
both the interactions. But the contribution from the spin-dependent 
part  comes out to be $\sim$ 2-3 orders of magnitude higher than that 
from spin-independent part. 
Henceforth for NEWAGE we show the event rates
with spin-dependent cross sections only.\\

For a detector with target material CS$_2$, the spin-independent directional recoil 
rate (as given in Eq.\ \ref{eqdedomega}) can be written as \citep{GONDOLO1}
\begin{eqnarray}
\frac{dR}{d\cos\theta \ d\phi} 
&=&
1.306 \times 10^{-3} \int dE\left(\frac{\rho}{0.3 {\rm GeV}}\right)
\left(\frac{\sigma}{10^{-44} {\rm cm^2}}\right) \nonumber\\
&& \times \frac{f_{\rm eff}^{\rm SI}(E,\v{w})}{4\pi\mu_p^2 m} 
\left[\frac{\rm events}{\rm kg \cdot day\cdot sr} \right] .
\label{eq:cs2}
\end{eqnarray}
In the above, $\sigma$ is the WIMP-proton cross section, 
$\mu_p$ is the WIMP-proton
reduced mass in GeV and $f_{\rm eff}^{\rm SI} (E,\v{w})$ is given by 
\begin{eqnarray}
f_{\rm eff}^{\rm SI} (E,\v{w}) &=& C_{\rm C}A_{\rm C}^2 \hat{f}_{\rm C}(v_C,\v{w})
{\cal F}_{\rm C} + C_{\rm S}A_{\rm S}^2 \hat{f}_{\rm S}(v_S,\v{w})  {\cal F}_{\rm S}.
\label{eqa:26}
\end{eqnarray}
The function $\hat{f}_{\rm C,\rm S}(v_{C,S},\v{w})$ 
can be evaluated using 
Eq.\ \ref{eqmomspec}.
${\cal F}_{\rm C}$, ${\cal F}_{\rm S}$ are squares of the nuclear form factors ($F(E)$)
given by $F(E) = \exp(E/2E_0)$ where $E$ is the energy transferred from  WIMP to 
the target nucleus, $E_0 = 3/(2m_N R^2_0)$, $m_N$ being
the nuclear mass and $R_0$ being 
the radius of the nucleus \citep{jungman}.
The form factor is more accurately given as \citep{jungman,engel}
$F(E) = \left[\frac{3j_1(qR_1)}{qR_1}\right]^2 \exp(-q^2 s^2)$, where the momentum 
transferred $q = \sqrt{2m_N E}$, $R_1 = (R^2 - 5s^2)^{1/2}$, 
$R \simeq 1.2  A^{1/3} {\rm fm}$, where $j_1$ 
is a spherical Bessel function, $s = 1 {\rm fm}$ and
$A$ is the mass number of target nucleus. 
The form factor depends on the distribution of mass and spin within the nucleus
and also on the type of the WIMP-nucleus interaction. 
For very small energy transfer $E$, relevant for our calculation, the form factor $\simeq 1$.
$A_{\rm C}$ and $A_{\rm S}$ are the mass number of
carbon and sulphur nuclei respectively.
$C_{\rm S}$ and $C_{\rm C}$ are the mass fractions of carbon 
and sulphur respectively and are given by
\begin{eqnarray}
C_{\rm C} &=& \frac{M_{\rm C}}{2M_{\rm S} +M_{\rm C} }  \nonumber \\
C_{\rm S} &=& \frac{M_{\rm S}}{2M_{\rm S} +M_{\rm C} }\,\,.  
\end{eqnarray}
In the above $M_{\rm C} (\approx 11.26 ~{\rm GeV})$ and $M_{\rm S} (\approx 30.04 ~{\rm GeV})$ are masses of carbon and sulphur
nuclei respectively. \\

The spin-dependent direction recoil rate 
for a detector with target material CF$_4$ is given by the same expression
as Eq.\ \ref{eq:cs2} but with $f_{\rm  eff}^{\rm SI}$ replaced by
$f_{\rm  eff}^{\rm SD}$ which is given by
\begin{eqnarray}
f_{\rm  eff}^{\rm SD}
&=&
\frac{4}{3} \ 0.647 \ C_F \hat{f}_F(v_F,\v{w})
\label{eqa:29}
\end{eqnarray}
where $C_F$ is the mass fraction of the fluorine nucleus given by
\begin{eqnarray}
C_F &=& \frac{4M_F}{4M_F + M_C}
\end{eqnarray}
$M_F$ being the mass ($\approx 17.84~ {\rm GeV}$) 
of the fluorine nucleus. The factor $(0.647)$ comes
from spin of fluorine nucleus and has been discussed in \citep{GONDOLO1}.\\
%

\begin{figure}[t]
\includegraphics[width=9cm, height=10cm, angle=0]{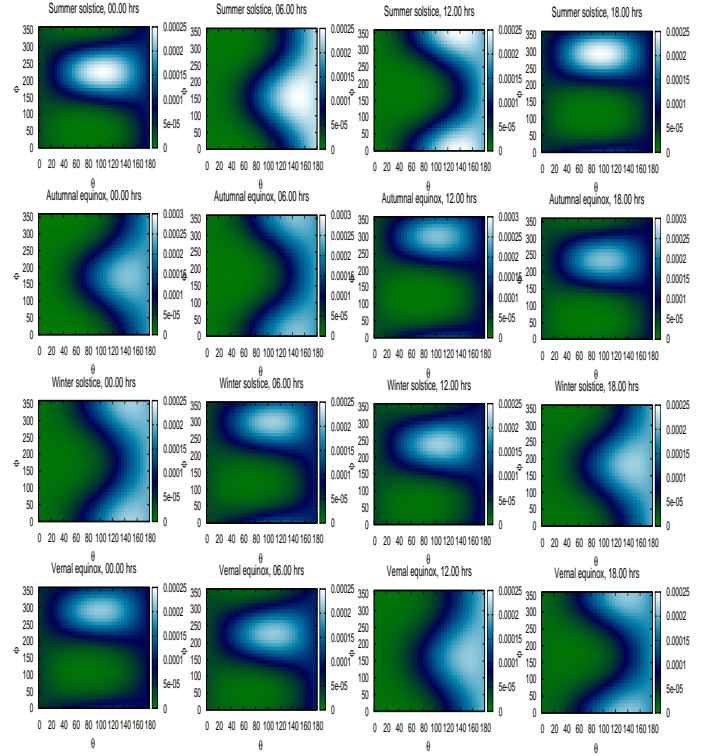}
\caption{\label{fig:colorcs2} These plots shows distribution of energy
integrated WIMP-nucleon spin-independent
event rates in CS$_2$ target  (DRIFT) detector
over $\theta$ and $\phi$ at different time of the year 
and at different instants of the day (described in the text) .
Color contour shows event rates in units of [events/kg/days/sr].
Plots are done with a cross section of $10^{-44}$ cm$^2$ and 
a dark matter mass of $60$ GeV.}
\end{figure}
\begin{figure}[t]
\includegraphics[width=9cm, height=10cm, angle=0]{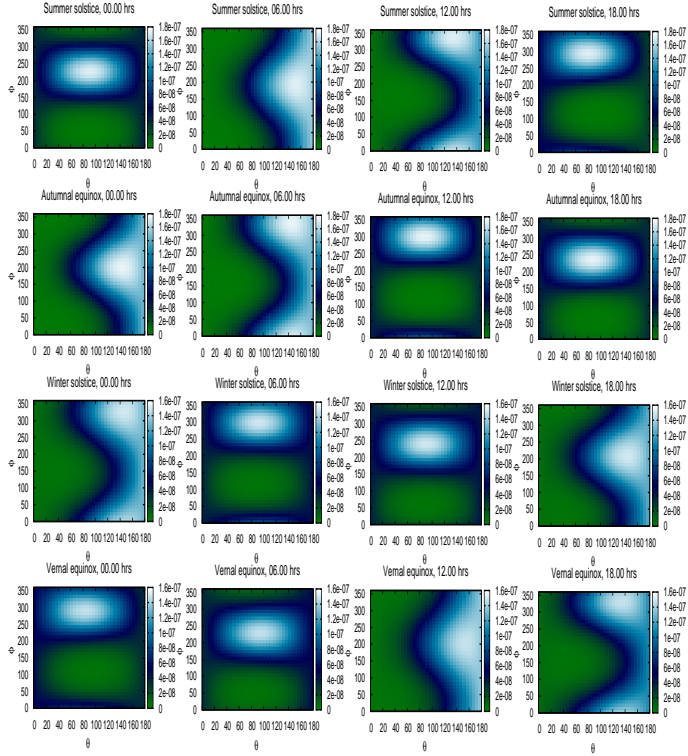}
\caption{\label{fig:colorcf4} Same as Fig.\ \ref{fig:colorcs2} but
for spin-dependent events in CF$_4$ target (NEWAGE)}
\end{figure}

\begin{figure}[h]
\begin{center} 
\includegraphics[width=8cm, height=8cm, angle=0]{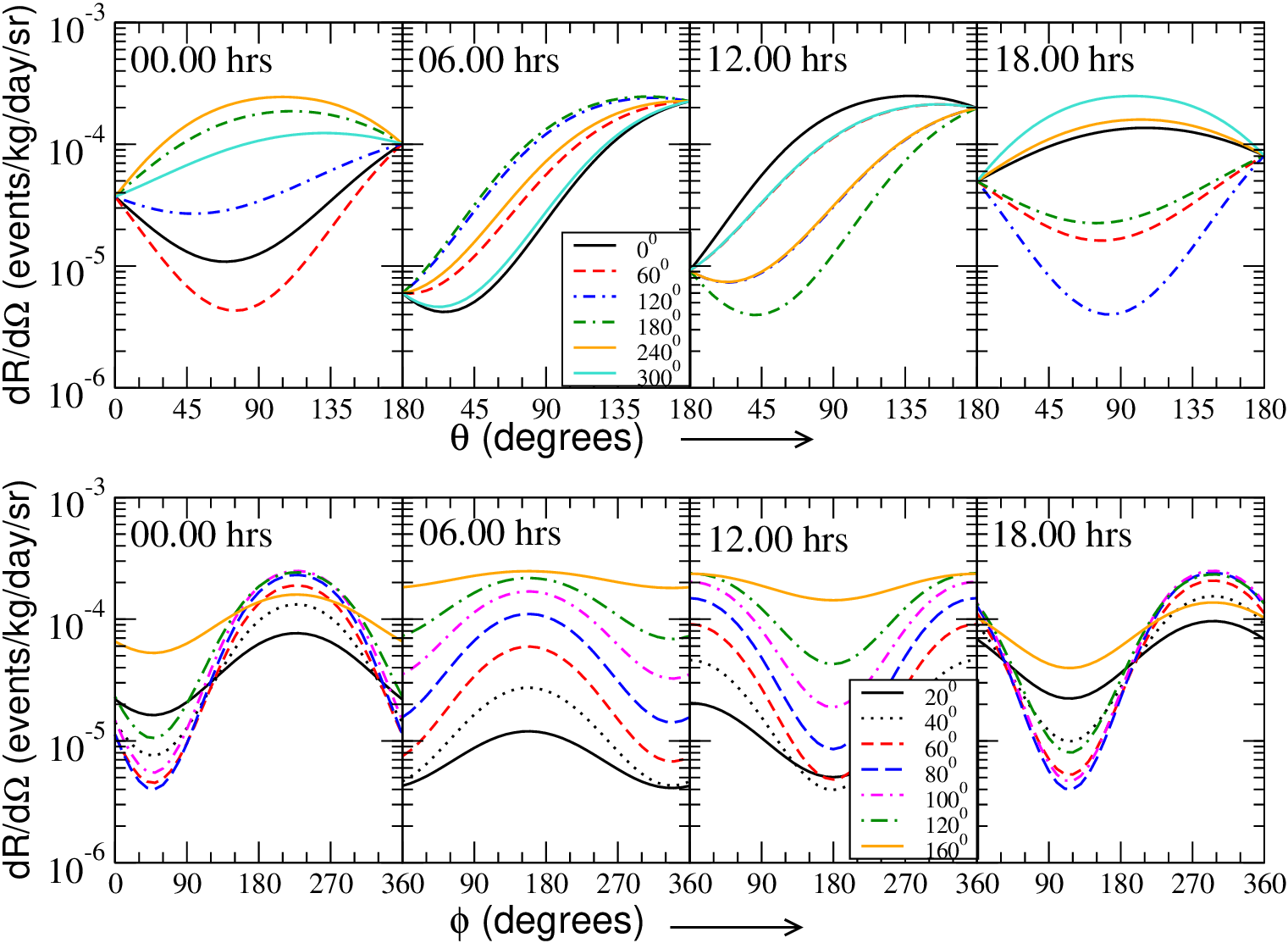}
\caption{\label{fig:cs2line} upper panel: $\frac{dR}{d\Omega}$ 
(spin-independent) vs $\phi$ at different
values of $\theta$  marked at the inset of 
$2^{nd}$ column. Lower panel: $\frac{dR}{d\Omega}$
(spin-independent) vs $\theta$ at different
values of $\phi$ marked at the inset of $3^{rd}$ column. 
We have chosen a WIMP-nucleon cross section of $10^{-44}$ cm$^2$.
dark matter mass of $60$ GeV. The target material is $CS_2$ (DRIFT
detector)}
\vskip 1cm
\includegraphics[width=8cm, height=8cm, angle=0]{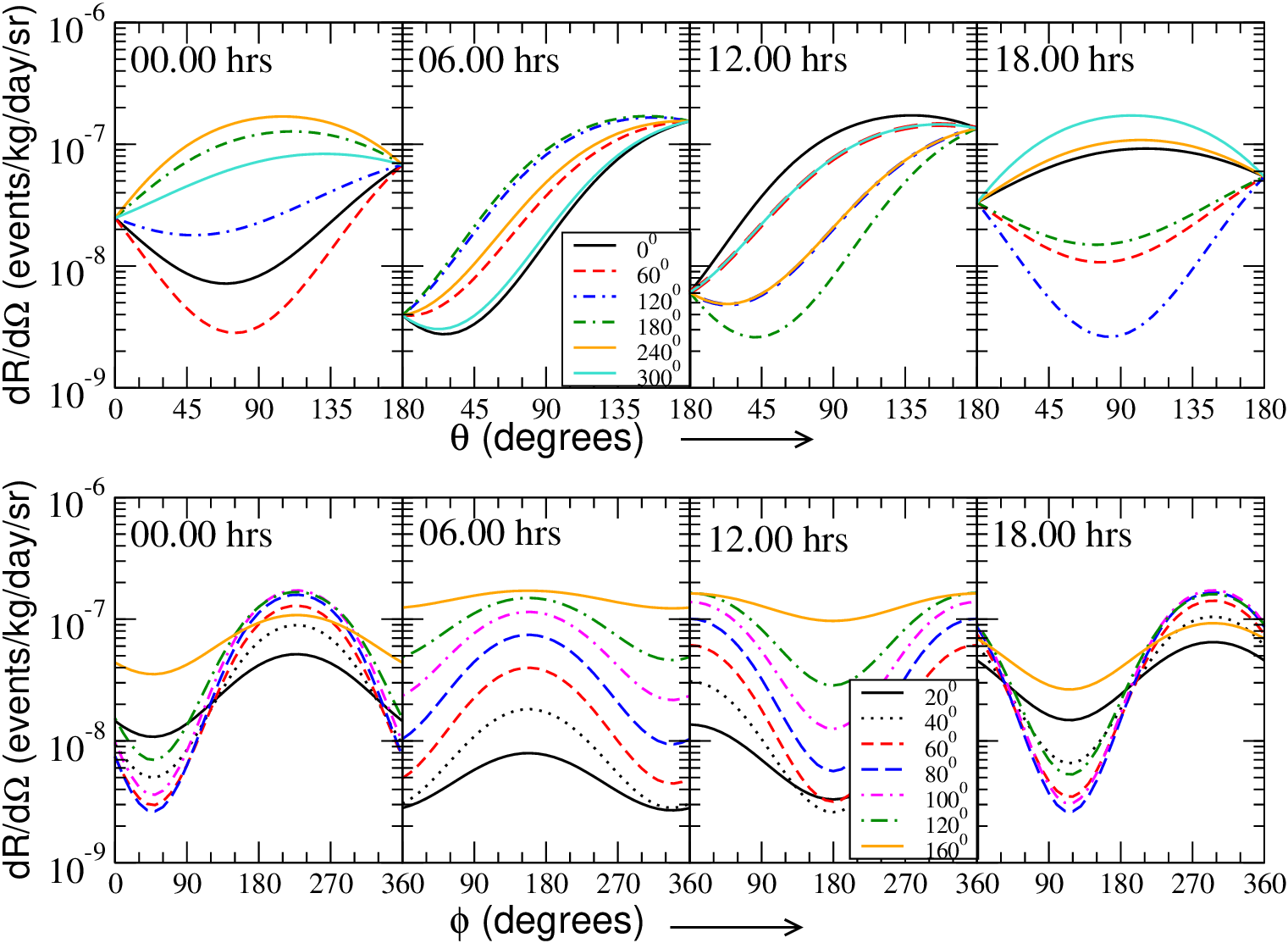}
\caption{\label{fig:cf4line}
Same as Fig.\ \ref{fig:cs2line} but for spin-dependent
interactions at a CF$_4$ in NEWAGE detector.}
\end{center}
\end{figure}
\begin{figure}[h]
\begin{center} 
\includegraphics[width=7cm, height=6cm, angle=0]{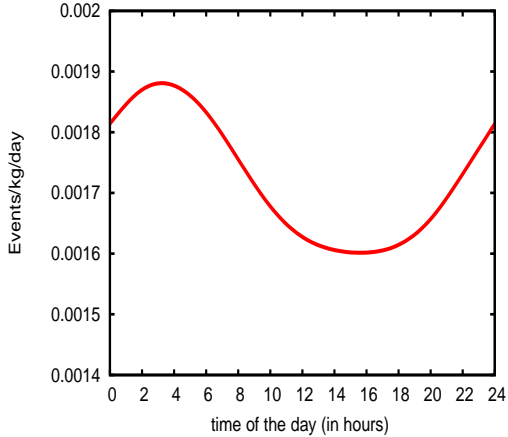}
\caption{\label{fig:san} Daily 
variation of $\theta$,$\phi$- integrated rates for DRIFT detector for WIMP mass of 60 GeV}
\end{center}
\end{figure}

The angular dependence of the spin-independent rates (number of events per day)
per kg of target material CS$_2$ of DRIFT detector are calculated using 
Eqs. (2-28) .
Similar calculations for spin-dependent rates with NEWAGE detector (CF$_4$)
are performed using Eqs. (2-26, 29-30).
The results for the spin-independent case (DRIFT) are shown in Fig.\ 
\ref{fig:colorcs2}.  The directional 
rates plotted here are calculated with a 
benchmark value of WIMP-nucleon cross section of $10^{-44}$ cm$^2$ and a 
WIMP mass of $60$ GeV. Each plot shows the variation of event rates with 
$\theta$ and $\phi$ as defined in Sec.\ \ref{sec:vw}. 
They are shown in Fig.\ \ref{fig:colorcs2} using a colour code symbol.
A colour corresponding to a point in the $\theta - \phi$ 
plane for any single plot in 
Fig.\ \ref{fig:colorcs2}, therefore indicates
the detection rate for the directionality given by the $\theta - \phi$ coordinates of that point.
The colour codes are chosen such that the increase
of detection rates can be represented by the change of colour from green to white 
in the order green -  deep blue - light blue - white.
The annual and diurnal 
variations of the directional dependence of the event rates observed in the 
detector are illustrated by presenting the plots for different values of the set 
$(T,t)$, where $T$ and $t$ represent a time of the year
and a time in a day respectively.
We show the results for $16$ representative sets of 
$(T,t)$  constructed out of four different times of the  year namely summer solstice,
autumnal equinox, winter solstice and vernal equinox and for four different times
of the day - 0.00 hrs (midnight), 6.00 hrs (6 A.M.), 
12.00 hrs (noon) and 18.00 hrs (6 P.M.). The plots of a given column represent the
annual variation of the directional rates observed at a particular time of 
the day. The plots of a given row, on the other hand, represent the diurnal 
variations of the directional rates observed at a particular time of the year. 
For example, from Fig.\ \ref{fig:colorcs2} we see that, for 
$(\theta,\phi)$ 
= (100$^\circ$ , 230$^\circ $) 
the directional rate is $5.5\times 10^{-6}$ 
events/kg/days/sr at 0.00 hours on the day of summer solstice 
(left corner plot of Fig.\ \ref{fig:colorcs2}). 
As one goes down along the first column of plots (annual variation) in 
Fig.\ \ref{fig:colorcs2} this rate changes to 
$4.5\times 10^{-5}$, $1.2\times 10^{-4}$, $2.9\times 10^{-5}$ in units of 
events/kg/days/sr for autumnal equinox, winter solstice and vernal equinox
respectively at 0.00 hours. The representative colour for the rates also changes 
accordingly. This rate also shows the variations in different times of a day
for the chosen value of $\theta,\phi$ as one moves row-wise from the 
left corner plot of Fig.\ \ref{fig:colorcs2}. In this particular example the rate
varies from $5.5\times 10^{-6}$ events/kg/days/sr to $6.5\times 10^{-5}$,
$1.5\times 10^{-4}$, $2.7\times 10^{-5 }$ (in units of 
events/kg/days/sr) that correspond to three different times of the day 
namely 6.00, 12.00 and 18.00 hours respectively on the day of summer solstice.
As is evident from Fig.\ \ref{fig:colorcs2},
these daily variations of the rates are also indicated by the change of 
representative colours. Similar observations can be made for plots 
in any other chosen rows or columns. In Fig.\ \ref{fig:colorcf4} 
we present the same plots as in Fig.\ \ref{fig:colorcs2} but for 
spin-dependent events in a CF$_4$ target
of NEWAGE detector. 
The pattern of variation of event rates with $\theta$ and $\phi$
differs in the two figures (Figs.\ \ref{fig:colorcs2} and \ref{fig:colorcf4}).
From Eqs.\ (\ref{eqa:26})  and (\ref{eqa:29}) we expect amplitude 
variation between the spin-dependent and spin-independent case.
Again from
Eq.\ (\ref{eq:25}) along with 
Eqs.\ (\ref{eqdiff}) and (\ref{eqmomspec}) one sees that
the directional rates also depend on the latitude ($\alpha$) 
of the detector location\footnote{
DRIFT is at a latitude $\alpha=54^\circ 33^\prime$N, for
NEWAGE $\alpha=36^\circ 25^\prime$N}.
This is reflected in Figs.\ \ref{fig:colorcs2} and \ref{fig:colorcf4}.
The diurnal variations of angular
dependence of the directional rates are also shown  for both
the cases of spin-independent and spin-dependent WIMP-nucleus
interactions in Fig.\ \ref{fig:cs2line} and Fig.\ \ref{fig:cf4line}
respectively for more clarity. In these two figures we have shown the variations of 
the directional event rates $dR/d(\cos\theta)d\phi$  with $\theta$ at
different values of $\phi$ (upper panel) and with $\phi$ at different
values of $\theta$ (lower panel). The plots are shown at four different times 
of a given day (the summer solstice). 
From Fig.\ \ref{fig:cs2line} it is evident that, for CS$_2$ detector,
 the range of directional variations (variations in $\theta$, $\phi$ ) of
the spin-independent rates lies between  $\sim (5\times 10^{-6} - 2\times
 10^{-4})$  events per day per unit solid angle per kg of detector mass.
The corresponding range for spin-dependent rates in a CF$_4$ detector is 
$(2\times 10^{-9} - 2\times 10^{-7})$ events/day/kg/sr 
( Fig.\ \ref{fig:cf4line} ). We also show in Fig.\ \ref{fig:san} the daily 
variation of $\theta$,$\phi$- integrated rates for DRIFT detector for WIMP 
mass of 60 GeV. The large daily variation is also apparent from this figure.
\\

\begin{figure}[t]
\begin{center}
\includegraphics[width=7cm, height=5cm, angle=0]{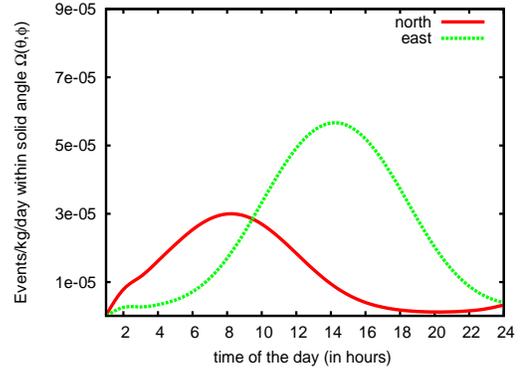}
\caption{\label{fig:alldir} The  number of nuclear recoils
in two solid angles with differently chosen directions  namely, north and east 
in the laboratory frame of reference as
a function of time of a day. The plots are shown at the day of summer solstice. }
\end{center}
\end{figure}

It is also evident from Figs.\ \ref{fig:colorcs2} - \ref{fig:cf4line} that calculated
daily and annual variations of detected number of events within a given solid angle
is sensitive to the chosen direction about which the solid angle is  considered.
For example, event number calculated for DRIFT 
around the direction specified (in the Laboratory frame) 
by $\theta=90^\circ$ and $\phi=300^\circ$ has a larger
variation over a day and also over the year. This can be understood from  Fig.\ \ref{fig:colorcs2}
as the representative colour associated with the point exhibits  a span of variations
ranging from green to white over different times of a day (or a year). On the contrary,
number of events calculated around the direction specified 
by $\theta=20^\circ$ and $\phi=25^\circ$ shows mild daily and annual variations.
This is realised from  Fig.\ \ref{fig:colorcs2} since over  different times of a day (or a year)
the point lies mostly in the green region of the $\theta-\phi$ plane.
The daily variations of direct detection rates for three chosen directions of recoil tracks
are shown in Fig.\ \ref{fig:alldir} for illustrative purposes. In  Fig.\ \ref{fig:alldir} we plot the
calculated number of events  within a solid angle considered about two different chosen 
directions namely north and east  
as a   function of different times of a sidereal day.
In this case we have chosen this day to be the day of summer solstice. The solid angles are so 
chosen that the semi-vertical angle of the cone around the chosen 
direction is 15$^\circ$.  It reveals larger daily variations along east 
direction in 
comparison to the northern direction for this case.  
Experiments dedicated to  directional dark matter detection 
may therefore look for
some preferential directions as obtained from such theoretical studies  for
the daily variations calculated in the Laboratory frame of reference,
along which the substantial daily variation is expected.\\

\begin{figure}[t]
\begin{center}
\includegraphics[width=6cm, height=6cm, angle=0]{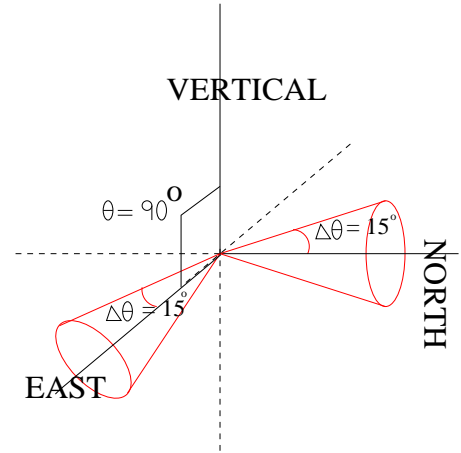}
\caption{\label{fig:cone} Schematic diagram to show the choice of solid angles as described in the text.}
\end{center}
\end{figure}

The diurnal variations  in the directional distribution of WIMP signals
can be better illustrated by 
studying the daily fluctuations of a dimensionless quantity 
defined as the ratio of directional recoil rate in a 
direction $\v{w}_1(\theta_1,\phi_1)$ to that in a 
direction $\v{w}_2(\theta_2,\phi_2)$.
In a practical scenario this can be estimated by considering ratio of number of events
for two solid angles $\Omega_1$ and $\Omega_2$ around the two specified directions ($\v{w}_1$ and $\v{w}_2$) which can be defined as
\begin{eqnarray}
r[\v{w}_1;\v{w}_2] &=& \frac{\int_{\Omega_1} \frac{dR}{d\Omega} d\Omega}
{\int_{\Omega_2} \frac{dR}{d\Omega} d\Omega} \nonumber\\
&=& \frac{\int_{\phi_1-\Delta\phi}^{\phi_1+\Delta\phi}  \int_{\theta_1-\Delta\theta}^{\theta_1+\Delta\theta} d\phi d\theta \sin\theta \frac{dR}{d\Omega}(\theta, \phi)}
{\int_{\phi_2-\Delta\phi}^{\phi_2+\Delta\phi} \int_{\theta_2-\Delta\theta}^{\theta_2+\Delta\theta} d\phi d\theta \sin\theta \frac{dR}{d\Omega}(\theta, \phi)}
\label{eq:ratio}
\end{eqnarray}

\begin{figure}[h]
\begin{center}
\includegraphics[width=7cm, height=5cm, angle=0]{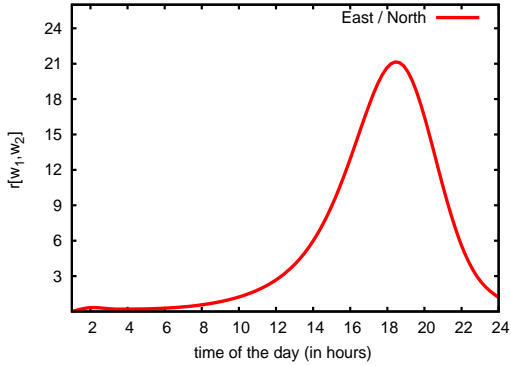}
\caption{\label{fig:ratio} The ratio of number of nuclear recoils
in two solid angles with  differently chosen directions 
in the laboratory frame of reference as
a function of time of a day. The plots are shown on the day of summer solstice.
We have presented the ratio for  a representative set :$(w_1,w_2)$: (east, north)}
\end{center}
\end{figure}

Such variations, estimated for the DRIFT detector, are shown in  
Fig.\ \ref{fig:ratio} where we have presented the daily variation of the
quantity $r[\v{w}_1,\v{w}_2]$ on the day of
summer solstice for a representative set:  $(\v{w}_1,\v{w}_2)$
chosen as \{east ($\theta = 90^\circ $, $\phi = 0^\circ$), north($\theta = 
90^\circ $, $\phi = 90^\circ$)\}.
with the semi vertical angle of the cones 
(that define the solid angles considered here)
taken to be $15^\circ$. 
The choice of the solid angles are illustrated in
Fig.\ \ref{fig:cone} where two choices of solid angles are shown - 
Our calculations show that the ratio of  detected events
corresponding to the eastward recoil direction to
that corresponding to the recoils along the north varies in the range 
$\sim 0.001 - 21$ over a sidereal day on the day of summer solstice.

\begin{figure}[t]
\begin{center}
\includegraphics[width=7cm, height=5cm, angle=0]{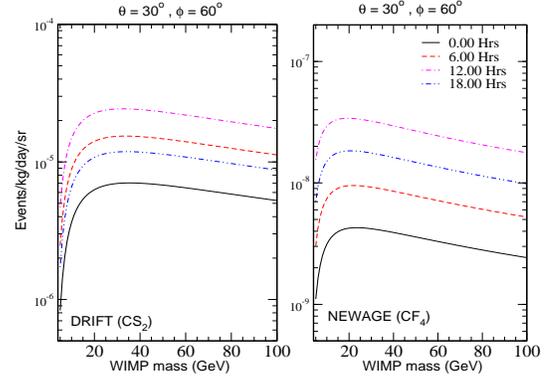}
\caption{\label{fig:drdm} Daily variation of detection
rates as a function of WIMP mass for DRIFT and NEWAGE detectors.}
\end{center}
\end{figure}

\begin{figure}[h]
\begin{center}
\includegraphics[width=7cm, height=5cm, angle=0]{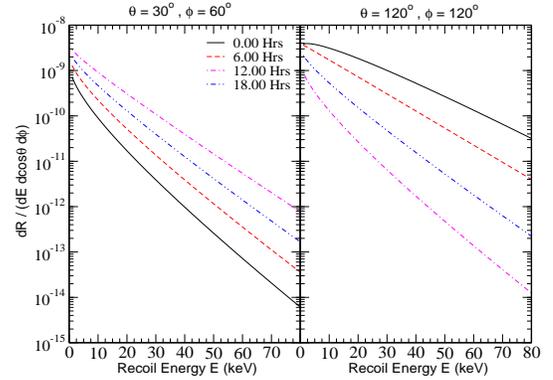}
\caption{\label{fig:drde} Daily variation of detection
rates as a function of recoil energy for DRIFT and NEWAGE detectors for a given WIMP mass of 60 GeV.}
\end{center}
\end{figure}

It may be of interest to investigate the daily variation of detection
rates as  functions of WIMP masses and recoil energies.
In Fig.\ \ref{fig:drdm} 
we show the energy integrated rates for different WIMP masses
at different time of the day. These rates are plotted for two detectors
- DRIFT and NEWAGE - considered here. The overall yield for each of the 
detectors is maximum at  WIMP masses around 20 GeV. The daily variation
also dominates for  WIMP masses at around 20 GeV for both the detectors
considered here. 
We also calculate the daily variation of rates with recoil energy for a fixed 
WIMP mass of 60 GeV and the results are shown in Fig.\ \ref{fig:drde}. 
From  Fig.\ \ref{fig:drde} one sees that the overall yield decreases with
the increase in recoil energy although the daily variation is more prominent 
for higher recoil energies.
\begin{figure}[h]
\begin{center}
\includegraphics[width=7cm, height=8cm, angle=270]{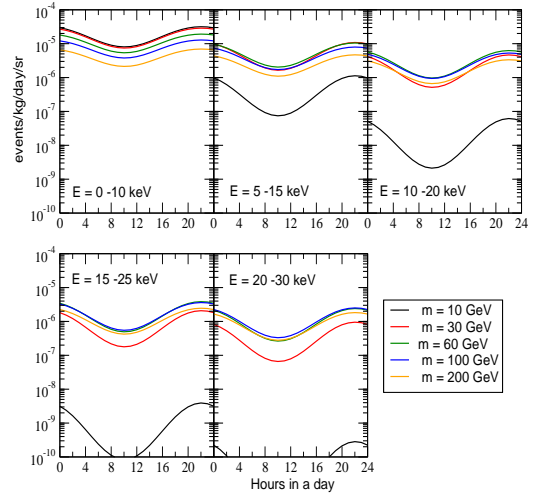}
\caption{\label{fig:em} Daily variation of detection
rates as a function of recoil energy and WIMP mass for DRIFT detector.}
\end{center}
\end{figure}
We also investigate the sensitivity of mass ranges to the daily
variations for different ranges of recoil energies. For demonstrative
purposes we show the results for DRIFT detector only. 
In  Fig.\ \ref{fig:em} the results are plotted for five fixed values 
of WIMP masses ranging from 10 GeV - 200 GeV. 
All the rates shown in this figure are integrated rates over 
$\theta$ and $\phi$ and over the considered energy ranges.
The  daily variations are investigated for
five recoil energy ranges: 0-10 keV, 5-15 keV, 10-20 keV, 15-25 keV, 
20-30 keV. Beyond 30 keV the rate falls down to very low value.
It is evident from Fig.\  \ref{fig:em} that the overall rate dominates for lower
recoil energy range (0-10 keV) all along the WIMP mass range considered.
The nature of daily variations of the detection rates are also similar for the
whole  mass range in the recoil energy band of 0 - 10 keV. 
It is also observed from Fig.\  \ref{fig:em} that for 
higher recoil energy ranges, the yield for lower WIMP masses gradually
decreases but the magnitude of daily variation increases. It appears
from the present calculations that for the particular detector considered,
lower recoil energy range is more suited for investigating 
dark matter if the WIMP masses are having a wide range of values 
(from very low to very high). But if WIMP masses are limited to only lower
values then the daily variation will be more prominent for higher recoil energy 
ranges.\\

\textit{Diurnal variations of detection rates in alternative galactic halo models:} 
We also calculate the diurnal variations of the directional detection rates 
considering a different galactic halo model. Such consideration 
enables us to compare the results obtained from the standard halo model
with that from this new model. 
The actual experimental results when compared with such calculations
may throw light on the nature of dark matter halo. To this end, we consider
a model proposed by \citep{ddisc} which is a combination of
the usual standard halo model and a proposed dark disc that
corotates with the galactic disc. In the dark disc (DD) model it is
 assumed that the DD kinematics matches with that of  Milky Way's thick 
disc \citep{bruch}. This in turn, gives a rotation lag represented by 
$\v{V}_{\rm lag}$ with respect to $\v{V_{SG}}$ 
described earlier in Eq.\ \ref{eq:vform} along 
with a different velocity dispersion. 
A Maxwellian velocity distribution for dark matter 
is also considered for dark disc model \citep{bruch} with 
$\v{V_{SG}}$ replaced by $\v{V}_{\rm lag}$ and in this calculation  a 
dispersion of 50km/s is adopted following \citep{bruch}.
\begin{figure}[t]
\begin{center}
\includegraphics[width=7cm, height=5cm, angle=0]{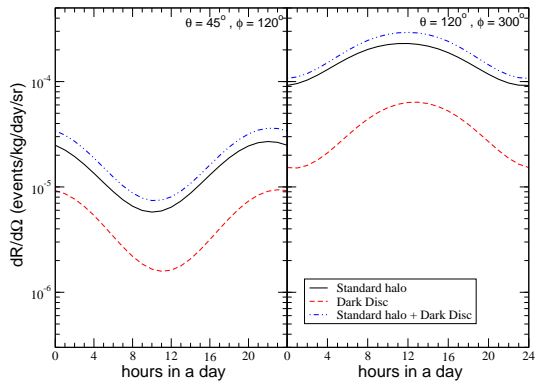}
\caption{\label{fig:dd} Comparison of diurnal variation of directional 
detection rates for dark disc model and standard halo model for dark matter for two sets of ($\theta,\phi$) on the day of summer solstice.}
\end{center}
\end{figure}
\begin{figure}[t]
\begin{center}
\includegraphics[width=7cm, height=5cm, angle=0]{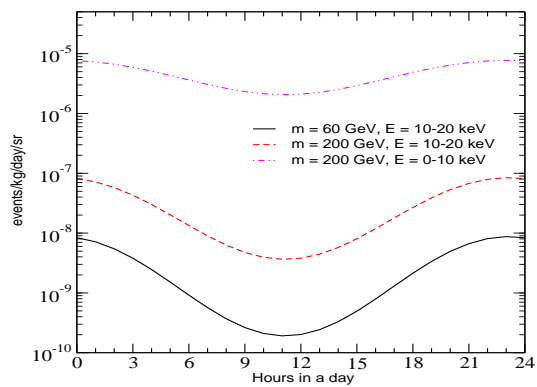}
\caption{\label{fig:ddonly}  Diurnal variation of directional 
detection rates for dark disc model with different dark matter mass and recoil energy range.
The plots are shown for a given 
$\theta(45^\circ),\phi(120^\circ)$ and on the day of summer solstice}
\end{center}
\end{figure}
In Fig.\ \ref{fig:dd} we show the daily variation of rate on a particular day
(summer solstice) for dark disc model and 
compared them with that of the standard halo 
model (SHM) for two sets of values of $\theta,\phi$. 
The rates from the combined model of DD and SHM  are also 
displayed in Fig.\ \ref{fig:dd}.
From Fig.\ \ref{fig:dd} one sees that the yield in case of 
standard halo model is more than that for dark disc
for the chosen value of dark matter mass.
The difference in the yield for the two cases 
varies when  observed in different $\theta,
\phi$ directions. The magnitude of daily variations also appears to differ 
marginally for the two models.
However we have checked that this difference between the two models
varies for different recoil energies and dark matter mass.
In Fig.\ \ref{fig:ddonly} the diurnal variation for two chosen dark matter
masses are shown for different recoil energy ranges considering only the 
dark disc model. One of the chosen dark matter mass is in the high mass range (200 GeV)
and the other one is chosen at 60 GeV as before. From Fig.\ \ref{fig:ddonly} one sees
that if only dark disc model is considered, the yield is much higher for the higher dark matter mass 
and low recoil energy range implying that the manifestation of the dark disc 
effect is more prominant for low recoil energy and large WIMP mass.

\section{Conclusion}
\label{sec:con}
Directional detection of dark matter in direct detection experiments
not only provides unambiguous signature of galactic WIMPs but also
serves as a potentially powerful probe to the structure and dynamics of
dark matter halo. The directional detection techniques involve
reconstruction and head-tail discrimination of tracks of nuclear 
recoil, induced by the scattering of WIMPs off the detector nuclei.
Directional description of the WIMP induced nuclear 
recoil rates in terms of angular coordinates defined and realised
in the laboratory frame of a detector exhibits temporal variation 
over a sidereal day due to diurnal motion of the earth. It would
also have an annual modulation owing to yearly variations of the
component of earth's orbital velocity along incoming WIMP direction.
Analytical details of the annual and diurnal variations are determined by
the dynamics and geometry of the sun-earth system and the location
(latitude) of the detector on the earth. 
Proper transformation of coordinate systems are also needed to obtain the 
results in terms of laboratory frame of reference. This is very crucial
for any detector based prediction of such directional modulations 
of dark matter 
signals. The directional rates
of WIMP signal events in directional detectors
sensitive to spin independent WIMP-nucleus interactions
(CS$_2$ target) are estimated for all possible recoil directions 
and they are found to
lie within the range  $\sim 10^{-6} - 10^{-4}$ per day per 
steradian per kg of target mass of the detector. 
A WIMP mass of $60$ GeV and WIMP-nucleon cross section
of $10^{-44}$cm$^2$ have been assumed for all the calculations.
Similar estimations for directional detectors with dominant
sensitivity to spin-dependent WIMP-nucleus interactions
gives the corresponding range in same units as  
$\sim 10^{-9} - 10^{-7}$. The diurnal variations in
the laboratory frame description of directional anisotropy of the 
WIMP induced nuclear recoils amount to a wide daily fluctuations
(ranging from 0.3 to 20) of the ratio of WIMP rates corresponding to 
two chosen orthogonal directions of nuclear recoils (east and north).\\

Energy integrated rates for different WIMP masses ranging from 0 - 100 
GeV show that the daily variation is most prominent for WIMP masses
around 20 GeV. The study (with WIMP mass of 60 GeV) reveals that
although the range of daily modulation increases with the increase in
recoil energies but the drop in overall yield for higher recoil energies
makes it difficult for a detector to register this modulation. The effect of
varied recoil energy ranges on the integrated 
(over $\theta$,$\phi$ and over the considered energy range)
yield and their daily modulations at DRIFT detector for a wide range of
WIMP masses (10-200 GeV) are also investigated. The results
show similar daily modulations for the mass range considered 
for lower recoil energy range (0 - 10 GeV). The overall yield is also
higher in this recoil energy range. Higher recoil energy ranges,
although results in the increase in magnitude of modulation for lower
WIMP masses but the overall yield drops down. Hence, if the detector is 
efficient and precise enough to detect the yield corresponding
to higher recoil energy ranges then for lower values of WIMP masses
this detector will observe considerable daily modulation of the
signal (as shown in Fig.\ \ref{fig:em}).\\

The results of the analysis are found to differ mostly in terms of
the total yield when another model for dark matter halo namely
the dark disc (where one considers a corotating disc of dark matter
with the galactic disc) model or DD model is considered. However, 
calculations using a combined model of DD and the usual SHM  
shows marginal increase in overall yield and also a variation of
nature of daily modulations for certain recoil directions.

{\bf Acknowledgements:} D.M. thanks J D Vergados for some discussions.\\


\end{document}